# Effect of acoustic streaming on tissue heating due to high-intensity focused ultrasound

**M A Solovchuk** [1], **T W H Sheu** [1,2], **M Thiriet**[3] and **W L Lin**[4]

[1]Department of Engineering Science and Ocean Engineering, National Taiwan University, No. 1, Sec. 4, Roosevelt Road, Taipei, Taiwan 10617, Republic of China
[2]Taida Institute of Mathematical Science (TIMS), National Taiwan University
[3]LJLL, University of Paris # 6, Paris, France
[4]Institute of Biomedical Engineering, National Taiwan University

E-mail: solovchuk@gmail.com, twhsheu@ntu.edu.tw

**Abstract.** The influences of blood vessels and focused location on temperature distribution during high-intensity focused ultrasound (HIFU) ablation of liver tumors is studied. A three-dimensional acoustics-thermal-fluid coupling model is employed to compute the temperature field in the hepatic cancerous region. The model is based on the linear Westervelt and bioheat equations as well as the nonlinear Navier-Stokes equations for the liver parenchyma and blood vessels. The effect of acoustic streaming is also taken into account in the present HIFU simulation study. Different blood vessel diameters and focal point locations were investigated. We found from this three-dimensional numerical study that in large blood vessels both the convective cooling and acoustic streaming can change considerably the temperature field and the thermal lesion near blood vessels. If the blood vessel is located within the beam width, both acoustic streaming and blood flow cooling effects should be taken into account. The predicted temperature difference for the cases considered with and without acoustic streaming effect is 54 % in regions close to the vessel wall. The necrosed volume can be reduced about 30 %, when taking into account acoustic streaming effect.

## 1. Introduction

Our liver is a highly perfused organ with the blood supplied by the hepatic artery and the portal vein (Thiriet 2008). Liver secretes bile, stores glycogen, distributes nutrient from the blood and gastrointestinal tract, and eliminates endo- and exogenous substrates and toxins. Liver is susceptible to primary and metastatic tumors. Statistically, liver cancer is now ranked as the second leading cause of death in Asia (Huang et al 2001). Hepatocellular carcinoma (HCC) is rapidly becoming the most common malignancy worldwide. Surgical resection has survival rate of only 25-30% in 5 years (Zhou 2011). A high risk of postoperative recurrence exists also for multi-focal malignancies. After repeated resections a poor success rate is expected.

HIFU therapy has been applied to ablate solid tumors in different areas of the body, including the pancreas, liver, prostate, breast, uterine fibroids, and soft-tissue sarcomas (Zhou 2011, Leslie and Kennedy 2007). In comparison with other conventional cancer treatment modalities such as surgery, radio- and chemo-therapy, HIFU has advantages to be non-invasive without ionization. Fewer complications after treatment were reported. However, lethal complications may develop if the vital blood vessels adjacent to the tumors are severely damaged.

When the liver tumor is proximal to large blood vessels, surgical treatment becomes complicated. Quite recently (Zhang et al 2009) it was first shown that HIFU can safely achieve a virtually complete necrosis of tumors close to major blood vessels. After a single session of HIFU treatment, the rate of complete necrosis was about 50%, which is not satisfactory at all. Lack of a complete response can be attributed to the large tumor size and cooling effect in large vessels.

Cooling from a large blood vessel may affect the treatment efficiency. A basic understanding of the factors that can influence the tissue necrosis volume is necessary to improve thermoablative therapy. In the past, temperature elevation in soft tissues was mostly modeled by the diffusion-type Pennes bioheat transfer equation, in which the heat source is produced by the incident acoustic wave and heat sink owing to the perfusion in capillaries (Pennes 1948). The amount of dissipated heat can be estimated by averaging the effect of blood perfusion over all tissues. Homogenization assumption is probably no longer valid when modeling the temperature elevation in the regions containing sufficiently large vessels, inside which the blood is flowing.

Both biologically relevant convective cooling in large blood vessels and perfusion cooling in the microvasculature need to be taken into account altogether. Inclusion of these two possible cooling means will greatly increase modeling complexity since the equations of motion for blood flow need to be solved together with the divergence-free velocity constraint equation. Kolios et al (1996) and Curra et al (2000) studied the influence of blood vessel on the lesion size. The temperature elevation during focused ultrasound surgery was computed by two coupled bioheat equations, one for the tissue domain and the other for the blood domain. The blood vessel was on the acoustic axis. They carried out a 2D finite difference calculation in cylindrical coordinate system. Hariharan et al (2007) presented 3D model to determine the influence of blood flow on the temperature distribution. Their model was constructed on the basis of acoustic and bioheat equations. However for real blood vessel geometry it is necessary to solve the nonlinear hemodynamic equations to get the velocity distribution.

Ultrasound propagating in a viscous fluid can induce an additional mass flow. This effect is known as the acoustic streaming. The 3D mathematical acoutics-thermal-fluid model presented in (Sheu et al 2011) takes the less explored acoustic streaming effect into account. Our aim is to show that the incident finite-amplitude ultrasound wave can affect the blood flow motion in large hepatic vessels and consequently the temperature distribution in tumor. Eckart (1948) theoretically studied the acoustic streaming profiles generated by uniform cylindrical ultrasound beam in an infinite circular tube. The beam width was much smaller than the tube diameter. He considered linear Navier-Stokes equations at the steady-state. However, for a case of high Reynolds numbers the nonlinear Navier-Stokes equations should be considered (Rudenko et al 1977, Nyborg 1998). Analytical results of earlier studies (Eckart 1948, Nyborg 1998) cannot be applied to predict the acoustic streaming velocities in blood vessels during HIFU therapy (Shi et al 2002). This study is aimed at investigating the acoustic streaming generated by high-intensity focused ultrasound in blood vessels with different diameters and for different focused beam orientations. Numerical result, which is computed from the heat transfer equation for large blood vessel and its surrounding tissue together with the nonlinear hemodynamic equations with an acoustic streaming effect, has not yet been reported for the simulation of HIFU tumor ablation. The whole physics remains nowadays poorly understood.

## 2. Methods

*2.1 Three-field coupling model*
*2.1.1 Acoustic equation for ultrasound propagation.* The Westervelt wave equation (Hamilton and Morfey 1998) given below for ultrasound pressure $p$ can be employed to model the finite-amplitude nonlinear wave propagation in a soft tissue, which is modeled as a thermo-viscous fluid

$$\nabla^2 p - \frac{1}{c_0^2}\frac{\partial^2 p}{\partial t^2} + \frac{\delta}{c_0^4}\frac{\partial^3 p}{\partial t^3} + \frac{\beta}{\rho_0 c_0^4}\frac{\partial^2 p^2}{\partial t^2} = 0 \qquad (1)$$

The first two terms describe the linear lossless wave propagating at a small-signal sound speed. The third loss term is due to thermal conductivity and fluid viscosity. The last term accounts for acoustic nonlinearity which can affect the thermal and mechanical changes within tissues (Bailey et al 2003). In soft tissues, which are assumed to be thermoviscous, the acoustic diffusivity $\delta$ accounts for the thermal and viscous losses in a fluid and is modeled by

$$\delta = \frac{2 c_0^3 \alpha}{\omega^2} \tag{2}$$

In the above, $\alpha$ denotes the acoustic absorption coefficient. In equation (1), $\beta = 1 + \frac{B}{2A}$ and $\omega (\equiv 2\pi f)$ are the nonlinearity coefficient and the angular frequency, respectively.

In the current HIFU simulation, we neglect the nonlinear acoustic effect, which will be the focus of our near future study, for simplifying the analysis within the linear context. In our study each small element $dS$ of the transducer surface is assumed to vibrate continuously with the same velocity $u = u_0 exp(i\omega t)$ in the direction normal to the surface. The resulting linear wave equation $\nabla^2 p - \frac{1}{c_0^2} \frac{\partial^2 p}{\partial t^2} + \frac{\delta}{c_0^4} \frac{\partial^3 p}{\partial t^3} = 0$ can be transformed to the diffraction integral for the velocity potential as follows (Pierce 1981, O'Neil 1949)

$$\psi_P = \iint_S \frac{U}{2\pi r} \exp[-(\alpha + ik)r] dS \tag{3}$$

In the above, $r$ is the distance from the source point on the transducer surface $dS$ to a field point $\bar{p}$, and $k$ is the wave number. The pressure amplitude at $\bar{p}$ can be calculated from the following expression

$$p_{\bar{P}} = ikc\rho_t \psi \tag{4}$$

where $\rho_t$ is the density of tissues, and $c$ is the speed of ultrasound in tissues. The ultrasound power deposition per unit volume is assumed to be proportional to the following local acoustic intensity $I$

$$q = 2\alpha I \tag{5}$$

where the intensity $I$ is defined as

$$I = \frac{p^2}{2\rho_t c} \tag{6}$$

Note that equations (3) and (4) consider only the effects of diffraction and attenuation without taking the effect of nonlinearity into account. Several studies (Hallaj and Cleveland 1999, Curra et al 2000, Bailey et al 2003, Hariharan et al 2007) showed that for the focal intensity in the range of 100 to 1000 W/cm$^2$ and the peak pressure smaller than 4 MPa, the physical complexity due to cavitation and nonlinearity field can be neglected with acceptable errors. If cavitation and acoustic nonlinearities are avoided, the lesion size and shape can be well prescribed from the thermal considerations. In the present study the acoustic energy emitted from the transducer is 80 Watt. The intensity generated at the focus is 327 W/cm$^2$ and the pressure at the focal point is 3.27 MPa. Therefore we don't consider here the effects of nonlinearity and cavitation.

*2.1.2 Thermal energy equation for tissue heating.* In this paper we will develop a biologically realistic thermal model by dividing the region of current interest into the region with tissue perfusion, which is attributed mainly to the capillary beds, and the capillary region containing blood vessels. In other words, the temperature field has been split into the domains for the perfused tissues and the flowing blood.

In a region free of large blood vessels, the diffusion-type Pennes bioheat equation (Pennes 1948) given below will be employed to model the transfer of heat in the perfused tissue region:

$$\rho_t c_t \frac{\partial T}{\partial t} = k_t \nabla^2 T - w_b c_b (T - T_\infty) + q \tag{7}$$

In the above energy equation, $\rho$, $c$, $k$ denote the density, specific heat, and thermal conductivity, respectively, with the subscripts $t$ and $b$ referring to the tissue and blood domains. The notation $T_\infty$ is denoted as the temperature at a location that is quite far from the heating focus. For $w_b$ ($\equiv 10$ $kg/m^3$s) shown in equation (7), it is known as the perfusion rate for the tissue cooling in capillary flows. It is noted that the above energy equation for $T$ is coupled with the linear acoustic equation (4) for the acoustic pressure through the power deposition term $q$ defined in equation (5).

In the region containing large vessels, within which blood flow can convect heat, the biologically relevant heat source, which is $q$, and the heat sink, which is $-\rho_b c_b \mathbf{u} \cdot \nabla T$, will be added to the conventional diffusion-type heat equation. The resulting model equation avoids a possible high recurrence

$$\rho_b c_b \frac{\partial T}{\partial t} = k_b \nabla^2 T - \rho_b c_b \mathbf{u} \cdot \nabla T + q \tag{8}$$

where $\mathbf{u}$ is the blood flow velocity. Owing to the presence of velocity vector $\mathbf{u}$ in the energy equation, we know that a biologically proper model for HIFU simulation comprises a coupled system of thermal-fluid-acoustics nonlinear differential equations. The heat sink has association with the hydrodynamic equations that will be described in the following section. The heat source is governed by the acoustic field equation described in the previous section.

Thermal dose developed by Sapareto and Dewey (Sapareto and Dewey 1984) will be applied to provide us a quantitative relationship between the temperature and time for the heating of tissues and the extent of cell killing. In focused ultrasound surgery (the involved temperature is generally above $50\,°C$), the expression for the thermal dose (TD) can be written as:

$$TD = \int_{t_0}^{t_{final}} R^{(T-43)} dt \approx \sum_{t_0}^{t_{final}} R^{(T-43)} \Delta t \tag{9}$$

where $R=2$ for $T >= 43°C$, and $R=4$ for $37°C < T < 43°C$. The value of TD required for a total necrosis ranges from 25 to 240 min in biological tissues (Sheu et al 2011, Sapareto and Dewey 1984, Liu et al 2007). According to this relation, thermal dose resulting from the heating of the tissue to $43\,°C$ for 240 min is equivalent to that achieved by heating it to $56\,°C$ for one second.

*2.1.3 Acoustic streaming hydrodynamic equations.* When blood vessel is located within a beam width of high-intensity ultrasound beam, according to equation (8) the temperature in the tumor can be changed significantly depending on the blood flow velocity. High-intensity ultrasound beam can alter blood flow velocity; therefore we will include the effect of acoustic streaming in our mathematical model.

In this study we consider that the flow in large blood vessels is incompressible and laminar. The vector equation for modeling the blood flow motion, subject to the divergence free constraint equation $\nabla \cdot \mathbf{u} = 0$, in the presence of acoustic stress is as follows (Kamakura et al 1995).

$$\frac{\partial \mathbf{u}}{\partial t} + (\mathbf{u} \cdot \nabla)\mathbf{u} = \frac{\mu}{\rho}\nabla^2 \mathbf{u} - \frac{1}{\rho}\nabla \mathbf{P} + \frac{1}{\rho}\mathbf{F} \tag{10}$$

In the above, $\mathbf{P}$ is the static pressure, $\mu$ (=0.0035 kg/ms) the shear viscosity of blood flow, and $\rho$ the blood density. In equation (10), the force vector $\mathbf{F}$ acting on the blood fluid due to ultrasound is assumed to propagate along the acoustic axis $\mathbf{n}$. The resulting nonzero component of $\mathbf{F}$ along the direction $\mathbf{n}$ takes the following form (Nyborg 1998)

$$\mathbf{F} \cdot \mathbf{n} = \frac{2\alpha}{c_0} I \tag{11}$$

The acoustic intensity $I$ shown above has been defined in equation (6). Amongst the second-order physical effects, only the acoustic streaming will be taken into account.

*2.2. Description of the problem*

The single element HIFU transducer used in this study is spherically focused with an aperture of 12 cm and a focal length of 12 cm. This transducer presumably emits a beam of spherically-shaped ultrasound wave propagating towards the targeted tissue under the current investigation. The parameters used in the current simulation are listed in table 1 (Duck 1990, Sheu et al 2011).

Table 1. Acoustic and thermal properties for the liver tissue and blood.

| Tissue | $C_0$, m/s | $\rho$, kg/m$^3$ | c, J/kgK | k, W/mK | $\alpha$, Np/m |
|---|---|---|---|---|---|
| Liver | 1550 | 1055 | 3600 | 0.512 | 9 |
| Blood | 1540 | 1060 | 3770 | 0.53 | 1.5 |

Here we consider the case with the linear dependence of attenuation coefficient on frequency (Duck 1990). In this study, the ultrasound of 1.0 MHz insonation is incident from a location that is exterior of the liver tumor. Typically, the duration of energy delivery ranges from 5 to 12 s (Fischer et al 2010). The solid tumor was assumed to be exposed to an 8 s ultrasound. The acoustic energy is 80 Watt. The blood vessel schematic in figure 1 is parallel to the acoustic axis. In this study different blood vessel diameters ranging from 1.0 mm to 6.0 mm are considered. The blood viscosity is 0.0035 kg/(m s).

*2.3. Simulation details*

The three-dimensional problem is analyzed using the commercially available CFDRC (CFDRC Research, Hunstille, AL, USA) software. A detailed description of the solution procedures can be found in our previous article (Sheu et al 2011). For the calculation of temperature and velocity fields the computational domain has the dimensions of 6cm*4cm*4cm. It is sufficient for the considered treatment time. In the focal region (3mm*3mm*20mm) the refined grids were generated with a mesh size of 0.2mm*0.2mm*0.4mm. The number of grids used in this study was 71 815 in the blood vessel and 414 000 in the liver. Mesh independence was assessed by comparing the temperature distribution in the final working mesh with the temperature obtained in a refined mesh, which is generated by increasing the number of cells by 30%. In these two meshes, the temperatures differ from each other only by an amount less than 1 %.

*2.4. Comparison with the experimental data*

In order to verify the theoretical analysis, measurements were carried out in the following experimental setting. The acoustic source and hydrophone were immersed in filtered and deionized water that is contained in a 74.5-cm-long, 36-cm-wide, and 50-cm-high tank, which is open to the atmosphere. A three-dimensional computerized positioning system is used to move the transducer along the beam axis and orthogonal directions. A single-element transducer has a focal length of 120 mm, an aperture of 120 mm, and frequency of 1.0 MHz. The transducer was driven by a continuous wave.

In figure 2, the measured pressure profile is plotted against the axial distances (in the focal plane). The solid lines and open circles correspond to the prediction and measurement results, respectively. These results were obtained in water at $25°C$ using the chosen 0.4 mm hydrophone (Onda HNA-0400). The efficiency of the transducer was measured by means of the radiation force balance. The electric energy of the transducer is equal to 1.9 Watt. The measured acoustic pressures are normalized by the focal pressure of 0.72 MPa. Good agreement between the measured and numerical results can be seen. Since power deposition is quadratic in pressure, the deviation between the measurement and model analysis in the low amplitude regions is not deemed significant.

## 3. Results and discussion

### 3.1 Validation of numerical model
Our three-dimensional computational model for the acoustic streaming field was validated by comparing the results with those of Kamakura et al (1995). In figure 3(a) we present the streaming velocity profile as the function of time at the focal point. Acoustic streaming is generated by the transducer with a Gaussian amplitude distribution, radius of 1 cm, 5-cm focal length, frequency 5MHz, and maximum source pressure amplitude 30 kPa. The difference between our 3D-model result and the 2D-model result of Kamakura et al (figure 2 of Kamakura et al 1995) is less than 1%.

The present computational model was also validated by comparing our results with the experimental results of Huang et al (2004). This comparison was made for the temperature field in a uniform phantom. The results obtained using the present model are compared with the experimental data of Huang et al (figure 4 of Huang et al 2004) in figure 3(b). The simulation results include the thermocouple artifact heating. The acoustic frequency, peak focal pressure, and sonication time were 1.0 MHz, 1.11 MPa, and 1 second, respectively. The radius of transducer was 7 cm, and the focal length is 6.3 cm. Our results are in good agreement with the experimental data.

Then we compare the calculated temperature fields with the experimental results of Huang et al (figure 8 of Huang et al 2004) in a tissue phantom with 2.6 mm blood vessel. Our results presented in figure 3(c) have good agreement with the experimental data of Huang et al (figure 8 of Huang et al 2004) for the mean flow velocities of 0 and 1.87 cm/s within the estimated uncertainties in temperature measurements.

### 3.2 Acoustic streaming buildup
First we will study how the absorbed ultrasound energy can change the velocity distribution in blood vessel. It is interesting to see how the diameter of blood vessel can affect the acoustic streaming profile.

In figure 6 we present the $z$ velocity components at the cutting planes $y=0$ and $z=0.12$ for two focal point locations in the blood vessel: at the center of blood vessel and on the blood vessel wall. The initial velocity is equal to zero, the diameter of blood vessel is $d=3$ mm. Acoustic streaming velocity is induced by the absorbed ultrasound energy. We can see that focused ultrasound can induce acoustic streaming velocities up to 12 cm/s. The velocity gradient associated with the acoustic streaming motion is very high, especially, near the boundaries. The point with the maximum velocity is located in the postfocal region due to the mass flowing out of the focal region.

When the focal point is at the center of blood vessel we have the axially flowing blood in the center of the blood vessel and reverse flow in the boundary region. The total mass flow through any cross section is equal to zero. The velocity increases very fast near the focus. Within 0.2 second the flow becomes steady-state. The dependence of maximum velocity on the blood vessel diameter is presented in table 2. We found that the peak velocity increases with the increasing blood vessel diameter. For a blood vessel diameter smaller than 2 mm, when the focal point is on the blood vessel wall the peak velocity is larger than that for the case with the focal point at the center of blood vessel.

Table 2. Maximum streaming velocity computed in the blood vessel. A) focal point is at the center of blood vessel; B) focal point is on the blood vessel wall

| Diameter, mm | Maximum velocity, m/s (A) | Maximum velocity, m/s (B) |
|---|---|---|
| 1.0 | 0.002 | 0.009 |
| 1.4 | 0.008 | 0.026 |
| 2.0 | 0.028 | 0.060 |
| 3.0 | 0.10 | 0.11 |
| 6.0 | 0.22 | 0.15 |

In figure 6 the simulated velocity profiles *w* at the focal plane is presented for different distances between the focal point and the blood vessel center for the blood vessel diameters *d*=1.4 and 3 mm. The largest velocity magnitude for the blood vessel diameter *d*=3 mm is obtained for the case when the focal point is at *x*=0.5 mm. Increasing the blood vessel diameter causes an increase of acoustic streaming velocity and, consequently, an increase of blood flow cooling. In the next sections we will see that the effect of acoustic streaming on the temperature elevation becomes more pronounced for larger blood vessel diameters.

In figure 7 the simulated velocity profiles are presented for the cases with and without acoustic streaming for the blood vessel diameter *d*=3 mm. We study the blood flow velocities 0.016 m/s (corresponding to the velocity in vein with *d*=3 mm) and 0.13 m/s (artery) for *d*= 3 mm. When we take into account the acoustic streaming effect, blood flow becomes asymmetric. If the initial velocity is increased, the acoustic streaming becomes less important. For the presented cases we get a larger velocity gradient near the blood vessel boundary, when the acoustic streaming effect is taken into account. This will increase the blood flow cooling and decrease the temperature rise.

*3.3 Effect of convective cooling on the temperature distribution*

In this section we will study the temperature distribution in liver, when the distance between the focal point and blood vessel is 1.0 mm. The blood vessel is parallel to the acoustic axis. The vessel diameter is 3 mm. At inlet, the blood flow cross-sectional average velocities are set at 0.016 m/s and 0.13 m/s, that correspond to the velocities in veins and liver arteries with diameter 3 mm (Hand 1998). The velocity profile is assumed to be fully developed.

In figures 8, 9 we can see the computed temperature contours at *t*=8 second in the liver at the cutting planes *y*=0 and *z*=0.12 for the cases with and without blood flow. In these figures we don't show the simulated temperature that is higher than $56°C$, because as we mentioned before it is the threshold value for the tissue necrosis. For the case without blood flow an ellipsoidal shape of lesion with the dimension 11.7 mm*2.4 mm*2.4 mm is obtained. The blood flow tends to reduce the lesion size in both axial and radial directions. For the case with flow the lesion size is reduced from 2.4 to 2.0 mm in the radial direction. The distance between the lesion and blood vessel wall is 0.2 mm. This explains why tissues proximal to blood vessel remain viable. A layer of tissues proximal to the blood vessel receives a smaller thermal dose due to blood cooling. Some studies showed (Kolios et al 1996) that higher focal intensities and smaller exposure times can reduce the distance between the lesion and the blood vessel wall.

In figure 10 we can see the temperature profiles computed at *t*=8 s and *t*=20 s along the focal axis z and the radial axis x. At *t*=8 s the peak temperature is 78.7 °C at the focus in the absence of blood flow. If we take into account the blood flow, the maximal temperature becomes 76.2 °C. The simulated temperature profile becomes asymmetric in the presence of blood flow. There is a very fast temperature drop near the blood vessel wall. The maximal temperature along the radial axis shifts a length of 0.1 mm from the focal point at *t* =8 s. At *t*=20 s, the peak temperature at the focal point is 46.0 °C for the case considering without blood flow and 42.4 °C for the case with blood flow. At *t*=20 second, in the presence of blood flow the peak temperature shifts a length of 0.9 mm from the focal point along the radial axis. At time *t=20* second there is a large difference between the temperature distributions for the cases with and without blood flow. The effect of blood flow cooling increases with the increased treatment time. If destruction of all cells near the blood vessel boundary is necessary, a shorter sonication time with higher power deposition is suggested to minimize the viable region. We have computed the temperature for several distances between the focal point and blood vessel boundary. The calculated results show that large blood vessel can significantly change the temperature distribution in liver tumor when the distance between the focal point and blood vessel boundary is equal to several millimeters. Next we will study the effect of acoustic streaming on the velocity and temperature distributions in the investigated domain.

*3.4 Effect of acoustic streaming on the temperature distribution*
In figure 11 we present the temperature difference at $t$=8 s (exposure time = 8 s) for the cases investigated with and without inclusion of acoustic streaming. The distance between the focal point and blood vessel wall (the gap) is 0.5 mm, the diameter of blood vessel is 3 mm. The blood flow velocity is 0.016 m/s (vein). The largest temperature difference is found at the point close to the blood vessel wall and the magnitude is about 8 $^0$C (figure 11(c)). On the blood vessel wall (0.0015, 0, 0.12) the temperature rise is 20.2 $^0$C without acoustic streaming effect and 13.1 $^0$C with acoustic streaming effect. This means that if we don't take into account acoustic streaming the temperature is overestimated by 54 %. Inside the blood vessel the temperature rise for the cases investigated with and without inclusion of acoustic streaming can differ by three times. At the distance 1.5 mm from the blood vessel wall the difference between the cases investigated with and without acoustic streaming effect is 3%.

For the vein with the diameter $d$=1.4 mm (figure 12) the largest temperature difference for the cases with and without acoustic streaming is about 5 $^0$C. The blood flow velocity is 0.01 m/s. At the point on the blood vessel wall (0.0007, 0, 0.12) the temperature rise is 22.7 $^0$C without acoustic streaming effect and 18.5 $^0$C with acoustic streaming effect (23 % difference). The effect of acoustic streaming on the temperature elevation is more pronounced for the blood vessels with large diameter.

The predicted temperatures plotted as the function of initial velocity (m/s) for the cases with and without acoustic streaming effect at the focal point and on the blood vessel wall ($d$/2, 0, 0.12) are shown in figure 13 for two blood vessel diameters $d$=1.4 mm and $d$=3 mm. Inlet average velocities $u$=0.01 m/s and $u$=0.08 m/s correspond to the velocity in vein and artery with the diameter $d$=1.4 mm. For the case of inlet velocity $u$=0.01 m/s the temperature difference for the cases considered with and without acoustic streaming is 65 % for $d$=3 mm and 23 % for $d$=1.4 mm on the blood vessel wall. For the inlet velocity $u$=0.08 m/s (corresponds to the velocity in artery with $d$=1.4 mm) the temperature difference is 23 % and 6.5 % for the diameters $d$=3mm and $d$=1.4 mm correspondingly. At the focal point the temperature rise difference for the diameter $d$=3 mm for the cases considered with and without acoustic streaming is 10 % and 2.4 % for the inlet average inlet velocity $u$=0.016 m/s (vein) and 0.13 m/s (artery), correspondingly. For large diameters ($d$=3 mm) the effect of acoustic streaming is more pronounced. When the inlet average velocity is increased, the effect of acoustic streaming decreases.

*3.5 Effect of acoustic streaming on the lesion size*
In figure 14 we present the predicted lesion boundaries in liver for the cases with and without acoustic streaming effect for the blood flow velocity 0.016 m/s. The lesion size is calculated based on the thermal-dose threshold value of 240 min. When taking into account the acoustic streaming effect, the lesion size can be reduced by 22 %. When the distance between the focal point and the blood vessel boundary becomes smaller, the effect of acoustic streaming becomes more pronounced. The simulated results (figure 14) without inclusion of acoustic streaming effect show that tissue close to blood vessel can be completely ablated. However, if the effect of acoustic streaming is taken into account, a thin layer of tissue (0.09 mm) close to blood vessel remains viable. So the effect of acoustic streaming cannot be neglected in the treatment planning. Zhang et al (2009) suggested that HIFU combined with chemotherapy can be used to increase the efficiency of the treatment and prolong the survival.

The simulated ablated tumor volumes, which vary with the initial velocity, for the cases with and without acoustic streaming are presented in figure 15. The distances 0.5 mm and 0.7 mm between the blood vessel boundary and focal point are considered. The diameter of blood vessel is $d$=3mm. When we increase the velocity from 0.01 m/s to 0.04 m/s, the volume calculated without taking into account the acoustic streaming effect is decreased by an amount of 25 %. When increasing the velocity from 0.04 m/s to 0.08 m/s the difference of the volumes calculated without taking into account the acoustic streaming effect is only 12.5 %. A further increase of the velocity will decrease the ablated volume by only a small amount.

In figure 15 we can see that inclusion of acoustic streaming into the current analysis decreases the ablated tumor volume. At the velocity 0.01 m/s the difference of volumes for the cases with and without acoustic streaming is equal to 28% for the gap 0.5 mm and 14 % for the gap 0.7 mm. With the increasing initial velocity the difference will decrease accordingly. For the case with the velocity 0.13 m/s the difference of volumes is only 3 % and 1.4 % for the gaps of 0.5 mm and 0.7 mm, correspondingly. Increasing the gap from 0.5 mm to 0.7 mm results in an increase in lesion volume. For larger gaps the importance of acoustic streaming effect becomes smaller. For the gap of 0.9 mm the predicted difference of volumes for the cases with and without acoustic streaming is equal to 5% for the gap=0.5 mm and velocity $u$=0.016 m/s (vein). As we can see in figure 14, inclusion of acoustic streaming can considerably change the distance between the lesion and the blood vessel wall. The predicted results show that acoustic streaming can affect the lesion size, when the distance between the focal point and blood vessel wall is less than 1 mm. The transducer used in a simulation has a half-pressure amplitude width of 1.1 mm. So when the blood vessel is located within the half-pressure amplitude width, the acoustic streaming effect is important.

In figure 16 the simulated ablated tumor volumes as the function of inlet average velocity for the cases with and without acoustic streaming are presented for the diameter of blood vessel $d$=1.4 mm. The ablated volumes for $d$=1.4 mm are 24% to 32 % larger than the corresponding ablated volumes for the diameter of blood vessel $d$=3 mm. Comparing with the results for the diameter $d$=3 mm in figure 15 (a), we can see that for $d$=1.4 mm the effect of acoustic streaming becomes smaller.

## 4. Conclusion

We have proposed a three dimensional physical model to perform the current HIFU simulation. The proposed model takes into account the convective cooling in large blood vessel and the perfusion due to capillary flows. Convective cooling in large blood vessel was shown to be able to reduce the temperature near large blood vessel. Acoustic streaming was also included in the simulation model. For the diameter of blood vessel 6 mm the maximum acoustic streaming velocity is 22 cm/s. A decrease of the blood vessel diameter causes a decrease of acoustic streaming velocity (table 2). Due to the cooling effect enhanced by acoustic streaming, the temperature rise near the blood vessel boundary is much smaller than the value predicted without taking into account the acoustic streaming effect. The predicted temperature difference for the cases considered with and without acoustic streaming effect is 54 % close to the blood vessel wall for the diameter of blood vessel 3 mm. For the smaller diameter of blood vessel the effects of blood flow cooling and acoustic streaming on the temperature distribution become smaller (figure 13).

It was shown that acoustic streaming can affect the lesion size and the shape generated by ultrasound. This demonstrates the necessity of taking into account both of the convective cooling and acoustic streaming effects for a simulation involving a large blood vessel, when the tumor is proximal to the thermally significant blood vessel. Some major conclusions drawn from the current study are summarized below. Firstly, if the distance between the ultrasound beam and blood vessel is equal to several mm (2-3 mm), the convective cooling should be taken into account and the homogenization assumption becomes no longer acceptable. Secondly, acoustic streaming effect is important, when the distance between the focal point and blood vessel wall is less than 1 mm.

**Acknowledgement**
The authors would like to acknowledge the financial support from the National Science Council under the projects NSC 99-2628-M-002-005.

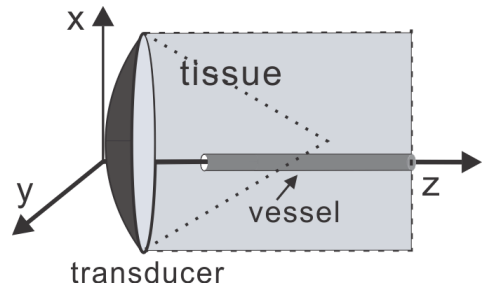

**Figure 1.** Schematic of the physical model. Blood vessel is parallel to the acoustics axis. The space enclosed by the dashed line and the transducer is the domain for conducting the current acoustic wave simulation.

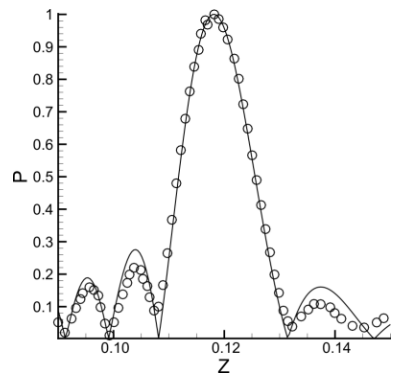

**Figure 2.** The measured (circle) and computed (solid line) normalized pressure profiles in water at 0.72 MPa focal pressure.

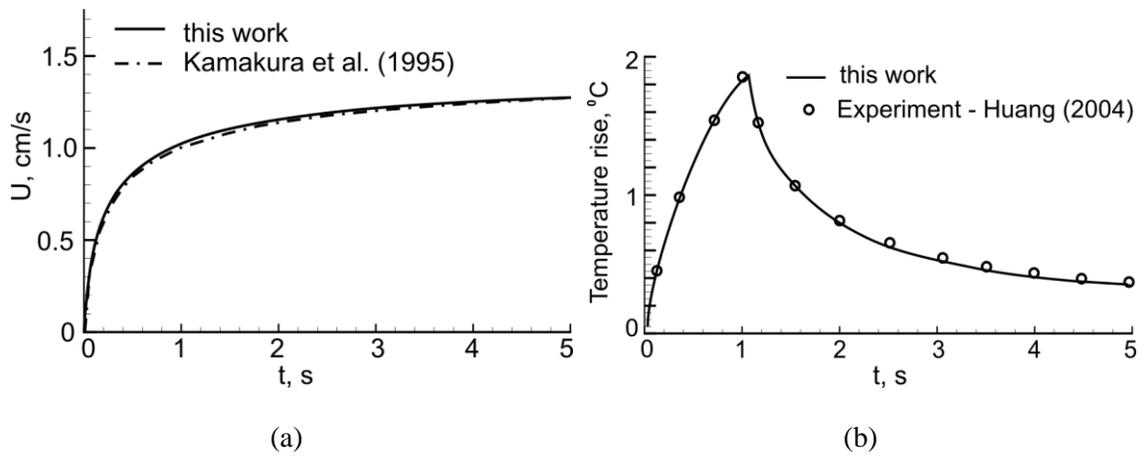

(a)                               (b)

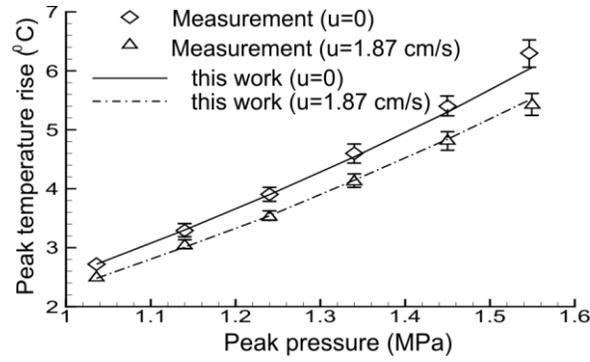

(c)

**Figure 3.** Validation of the computational model. (a) The currently predicted results are compared with the streaming velocity data of Kamakura et al (1995); (b) Comparison of the simulated focal temperature in a uniform phantom with the experimental data (Huang et al 2004); (c) Comparison of the simulated peak temperature rise at the focus (positioned in a phantom material and 0.4 mm from the vessel wall) with the experimental data (Huang et al 2004) as a function of pressure for different flow speeds in the vessel.

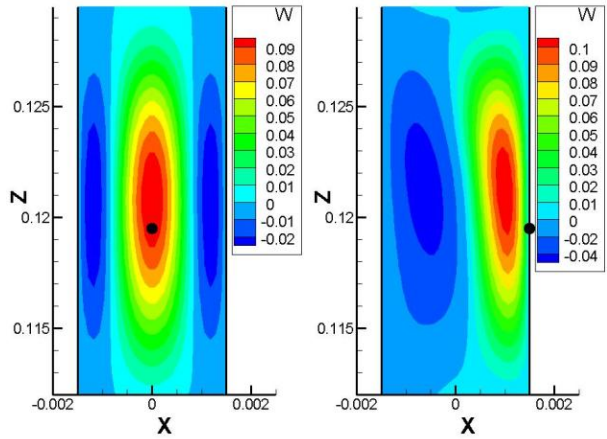

(a)  (b)

**Figure 4.** The simulated streaming profiles at the cutting plane $y = 0$ m without the externally applied flow (initial velocity=0) for two distances between the focal point (circle) and the vessel wall, $d=3$ mm. (a) focal point is at $x = 0$ m and $z = 0.12$ m in the center of blood vessel; (b) focal point is at $x = 0.0015$ m and $z=0.12$ m on the blood vessel wall.

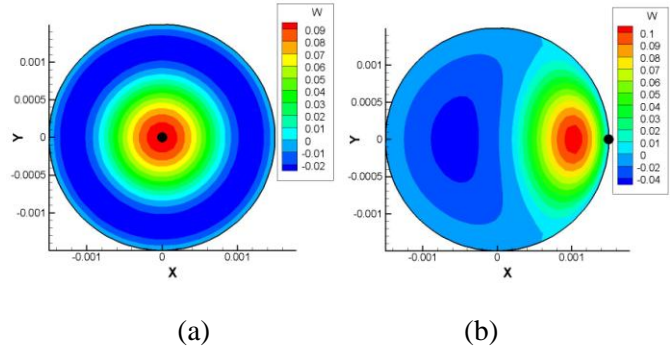

(a)            (b)

**Figure 5.** The simulated streaming profiles at the cutting plane $z = 0.12$ m without the externally applied flow (initial velocity=0) for two distances between the focal point and the vessel wall, $d$=3 mm. (a) focal point is at $x = 0$ m and $z = 0.12$ m in the center of blood vessel; (b) focal point is at $x$=0.0015 m and $z$=0.12 m on the blood vessel wall.

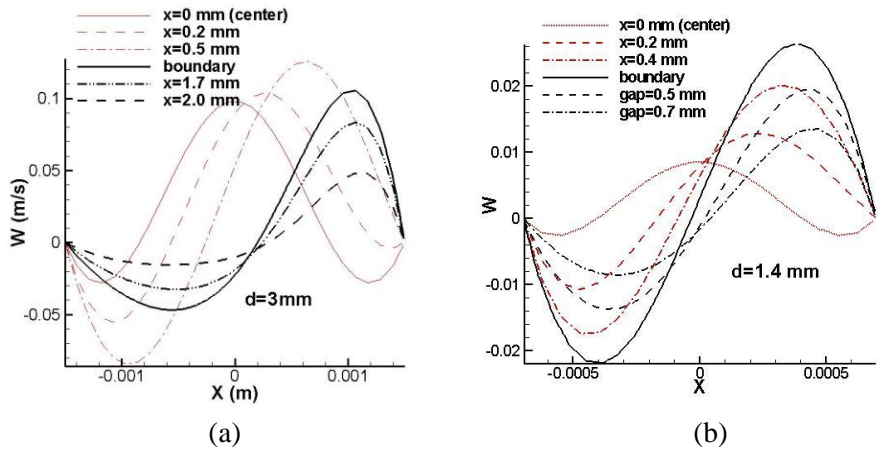

(a)            (b)

**Figure 6.** The simulated velocity profiles $w$ (x; y = 0; $z$=0.12 m) for different distances between the focal point and blood vessel center. (a) $d$=3 mm; (b) $d$=1.4 mm.

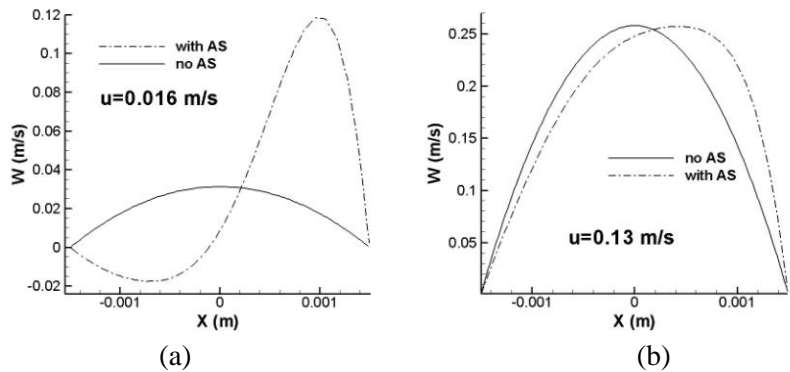

(a)            (b)

**Figure 7.** The simulated velocity profiles $w$ at the focal plane for the cases with and without acoustic streaming. Focal point is located on the vessel wall, ***d*=3 mm**. (a) inlet average velocity is 0.016 m/s (vein); (b) inlet average velocity is 0.13 m/s (artery).

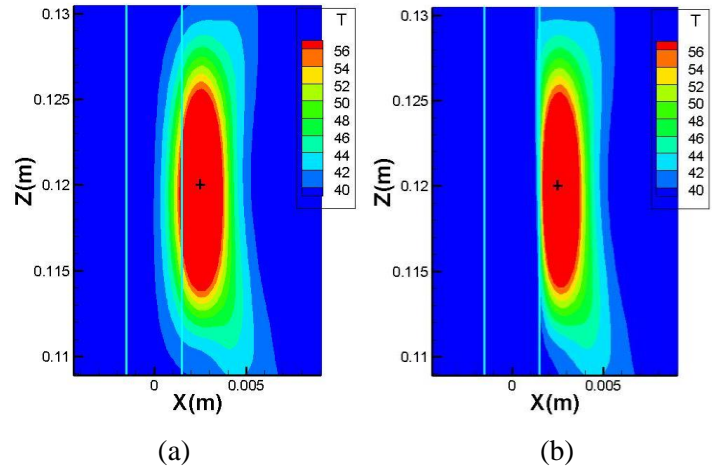

**Figure 8.** The predicted temperature contours at *t=8* s (just after sonication) in liver at the cutting plane *y=0* for the case investigated at f=1.0 MHz and 3.0 MPa pressure at the focal point (0.0025, 0, 0.12), *d*=3 mm, gap=1 mm. (a) without blood flow; (b) with blood flow, *u*=0.13 m/s.

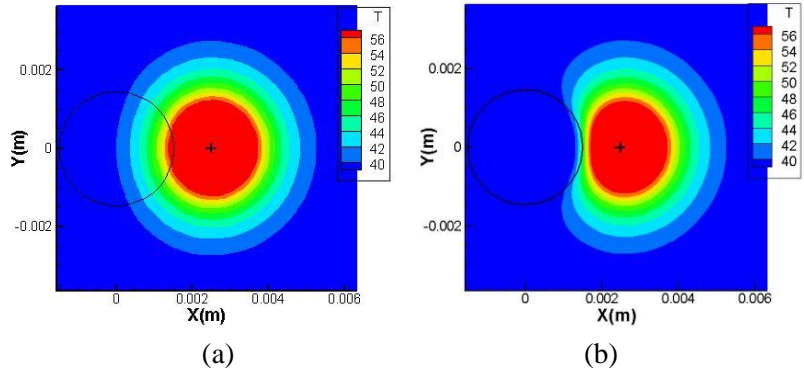

**Figure 9.** The predicted temperature contours at *t=8* s in liver at the cutting plane *z=0.12* m for the case investigated at *f*=1.0 MHz and 3.0 MPa pressure at the focal point, cross (+) denotes the location of focal point, *d*=3 mm, gap=1 mm. (a) without blood flow; (b) with blood flow, *u*=0.13 m/s.

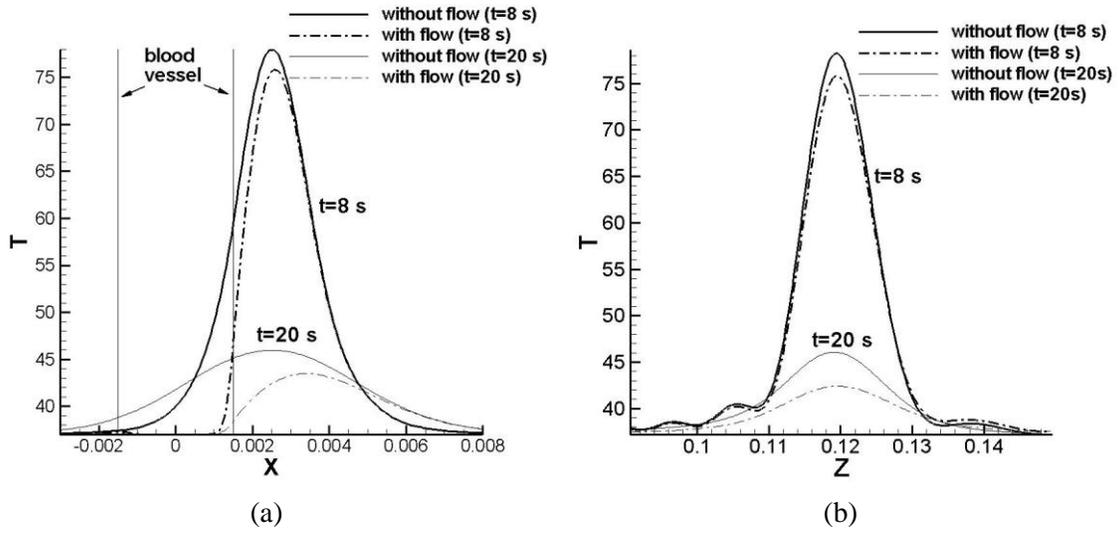

(a)            (b)

**Figure 10.** The predicted temperature distributions at *t=8 s* and *t=20* s along *x* and *z* directions for the cases with and without blood flow investigated at *f*=1.0 MHz.

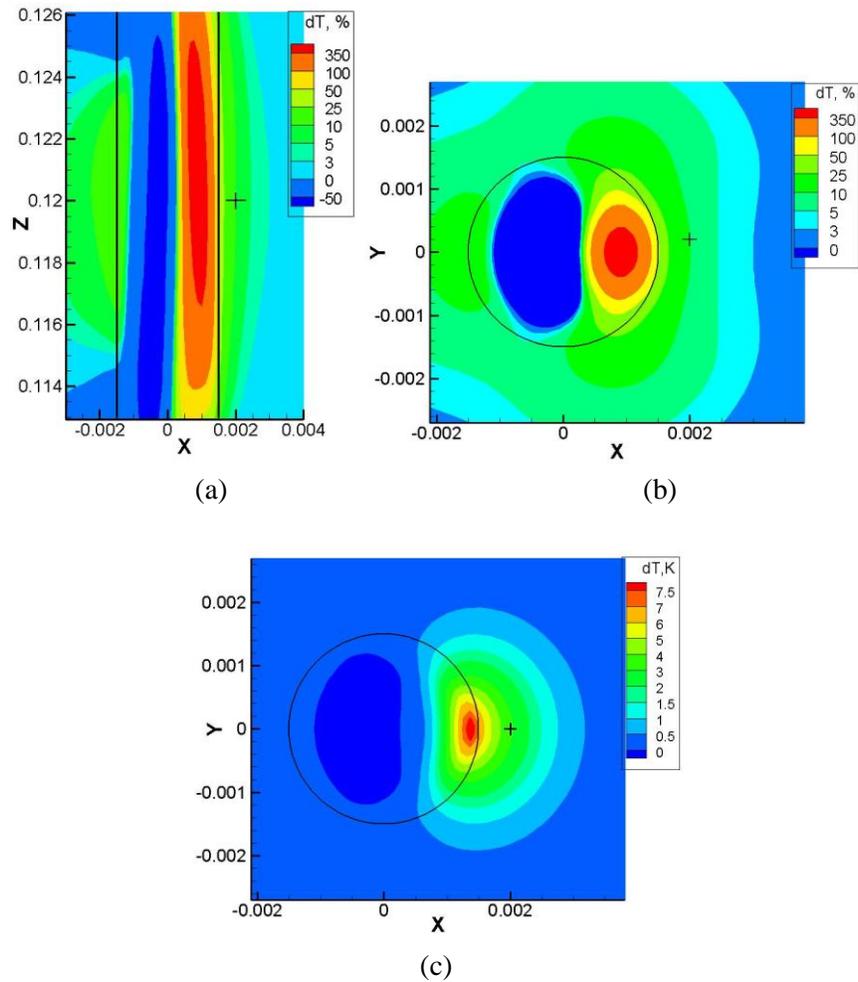

**Figure 11.** The temperature difference at *t*=8 s for the cases with and without acoustic streaming effect, gap=0.5 mm, ***d*=3 mm**, blood flow velocity is 0.016 m/s (vein). (a) the relative temperature difference at the cutting plane *y=0*; (b) the relative temperature difference at the cutting plane *z=0.12*; (c) the absolute temperature difference at the cutting plane *z=0.12*.

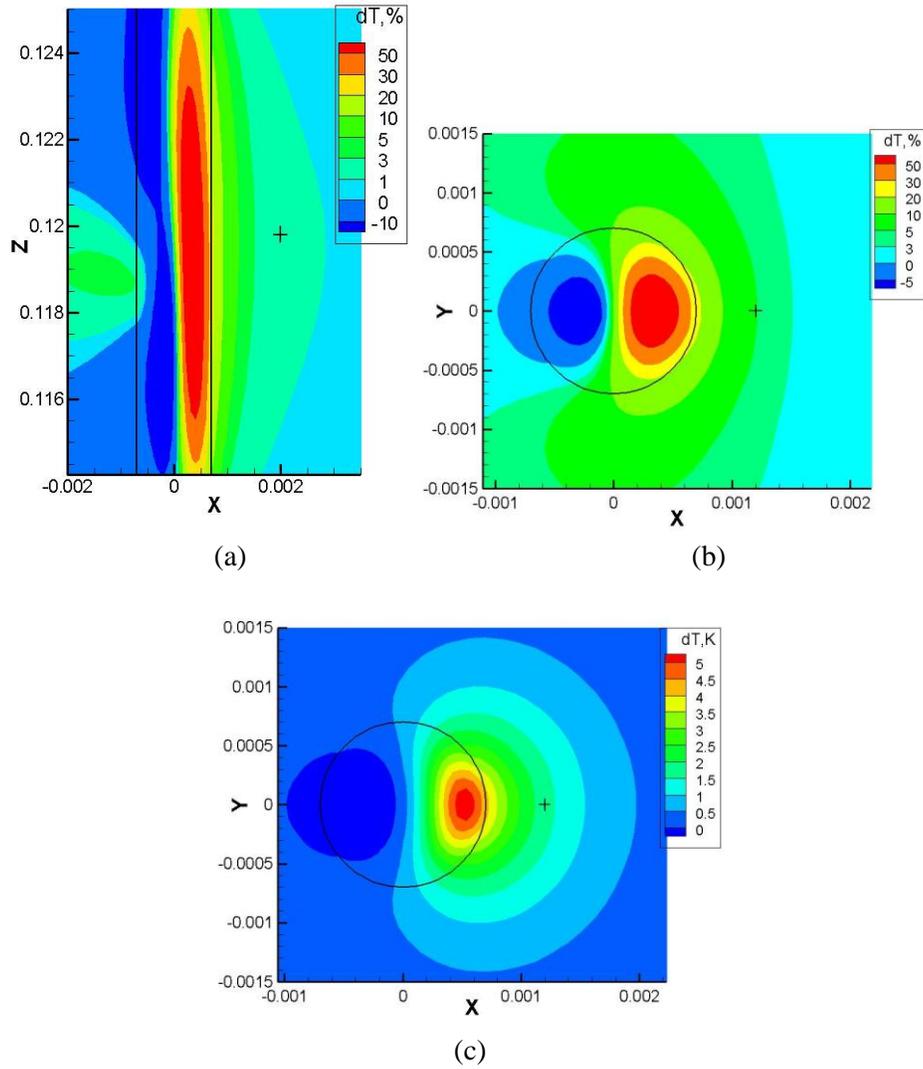

**Figure 12.** The relative temperature difference at *t*=8 s for the cases with and without acoustic streaming effect, gap=0.5 mm, ***d*=1.4 mm**, blood flow velocity is 0.01 m/s (vein). (a) the relative temperature difference at the cutting plane *y=0*; (b) the relative temperature difference at the cutting plane *z=0.12*; (c) the absolute temperature difference at the cutting plane *z=0.12*

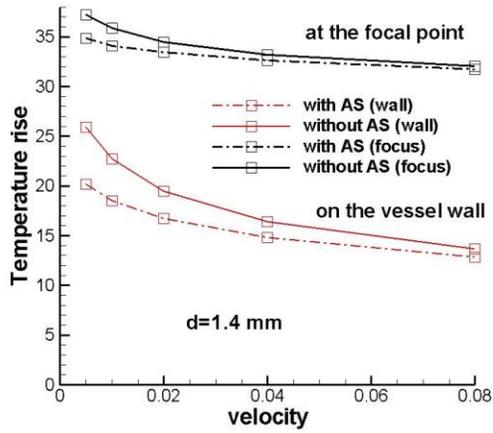
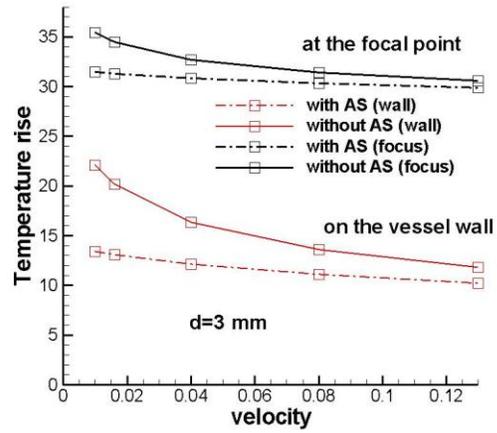

(a)                  (b)

**Figure 13.** The predicted temperatures as the function of inlet average velocity (m/s) for the cases with and without acoustic streaming effect at the focal point and on the blood vessel wall ($d/2$, 0, 0.12), $t = 8$ s, gap=0.5 mm. (a) $d$=1.4 mm; (b) $d$=3 mm.

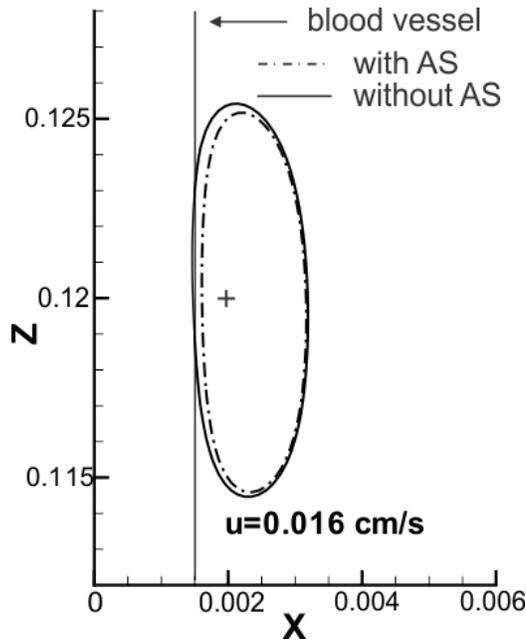
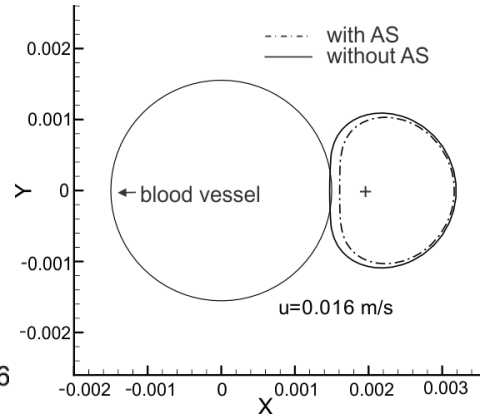

(a)                  (b)

**Figure 14.** The predicted lesion shapes in liver for the cases with and without acoustic streaming effect, $t$= 8 s, gap=0.5 mm, blood flow velocity is 0.016 m/s. (a) at the cutting plane *y=0*; (b) at the cutting plane *z=0.12*.

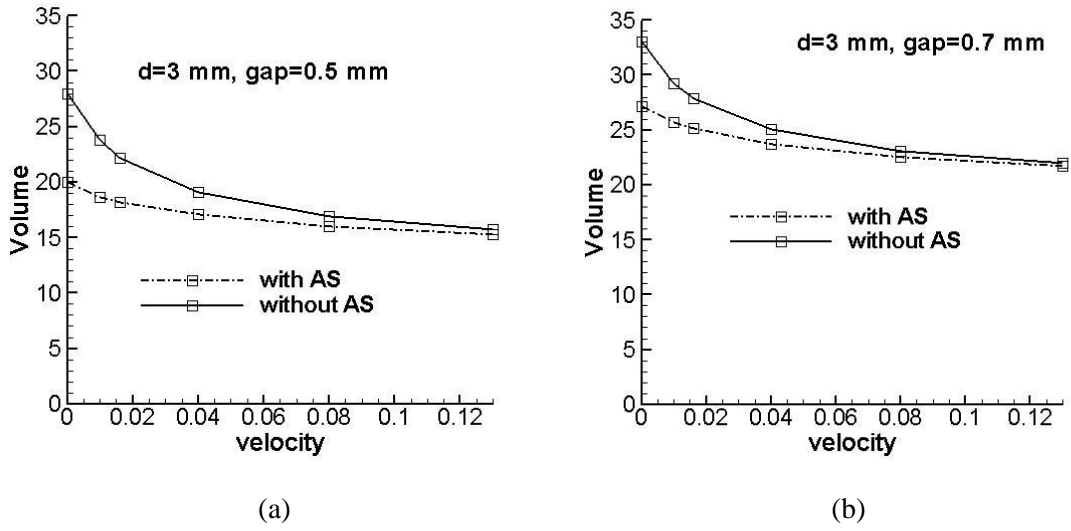

**Figure 15.** The simulated ablated tumor volumes (mm$^3$) as the function of inlet average velocity (m/s) for the cases with and without acoustic streaming, *d*=3mm. (a) gap=0.5 mm; (b) gap=0.7 mm.

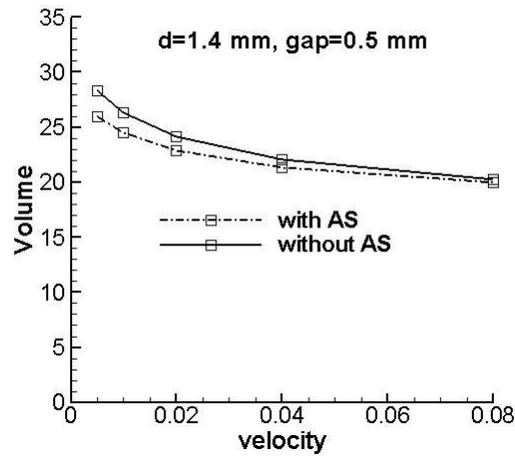

**Figure 16.** The simulated ablated tumor volumes (mm$^3$) as the function of inlet average velocity (m/s) for the cases with and without acoustic streaming, *d*=1.4 mm, gap=0.5 mm.